\documentclass [11pt,twoside]{article}

\usepackage{shrthnds}
\usepackage{cite}
\usepackage{graphicx}
\usepackage{color}
\usepackage{subfigure}
\usepackage{setspace}

\usepackage{multirow}

\usepackage[hang,small]{caption}

\usepackage{geometry}
\usepackage{fancyhdr}
\usepackage{titlesec,titletoc}
  \titleformat{\section}{\Large\sf\bfseries}{\thesection}{1em}{}
  \titleformat{\subsection}{\large\sf\bfseries}{\thesubsection}{1em}{}

\title{\sf\bfseries Scale Invariance as a
Solution to the Cosmological Constant Problem}

\author{\normalsize  Pankaj Jain$^{1}$\footnote{email: pkjain@iitk.ac.in}~,
 Subhadip Mitra$^{2}$\footnote{email: smitra@iopb.res.in}~,\\ Sukanta
Panda$^3$\footnote{email: sukanta@iiserbhopal.ac.in} and Naveen K. Singh$^1$\footnote{email: naveenks@iitk.ac.in}}
\date{}

\begin{document}
\maketitle
\vspace{-0.6cm}
\bc
{\small 1) Department of Physics, IIT Kanpur, Kanpur 208 016, India\\
2) Institute of Physics, Bhubaneswar 751 005, India\\
3) Indian Institute of Science Education and Research, Bhopal 462 023, India}
\ec

\centerline{\small\date{\today}}
\vspace{0.5cm}

\bc
\begin{minipage}{0.9\textwidth}\begin{spacing}{1}{\small {\bf Abstract:}
We show that scale invariance provides a solution to the
fine tuning problem of the cosmological constant.
We construct a generalization of the standard model of particle physics
which displays exact quantum scale invariance.
The matter action is invariant under global scale transformations in arbitrary
dimensions. However the gravitational action breaks scale invariance
explicitly. 
The scale symmetry is broken spontaneously in the matter
sector of the theory.
We show that the contribution to the vacuum energy and hence the
cosmological constant is identically zero from the matter sector
within the full quantum theory. However the gravitational sector may
give non-zero contributions to the
cosmological constant at loop orders.
 No fine
tuning may be required at loop orders since the matter sector gives zero
contribution to the cosmological constant.
We also show that we do not require full scale invariance in order
to constrain the vacuum energy from the matter sector. We only require
invariance under pseudoscale transformations. Using this idea and
motivated by the concept of unimodular gravity we propose an alternative
model. In this case also we show that matter sector gives exactly zero
contribution to the vacuum energy.

}\end{spacing}\end{minipage}\ec

\vspace{0.5cm}\begin{spacing}{1.1}

\section{Introduction}

The standard model of particle physics generically gives rise to a very 
large value of vacuum energy and hence suggests the presence of a cosmological constant 
many orders of magnitude larger than what is allowed by the observations. 
This has to be canceled by explicitly introducing a large cosmological constant term of opposite
sign, leading to a fine tuning problem
\cite{Weinberg,Peebles,Padmanabhan,Copeland,Carroll,Sahni,Ellis}.
Different approaches for solving the cosmological constant problem have been
considered by several authors
\cite{Weinberg,Aurilia,VanDer,Henneaux,Brown,Buchmuller,Buchmuller1,Henneaux89,Sorkin,Sahni1,Sundrum}.
We consider the idea of scale invariance. It has the potential to 
provide an answer to the fine tuning problem associated with the cosmological
constant. Scale invariance naturally rules out a cosmological constant \cite{JMS,Wetterich}.

In recent papers  \cite{JMS,AJS,JM09,JM10_1,JM10_2} we have studied a 
scale invariant generalization of the Standard Model of particle physics, 
including gravity \cite{ChengPRL}. The model contains no dimensionful 
parameter and displays local scale invariance -- an idea first proposed by
Weyl back in the twenties\cite{Weyl:1929} and later explored by many others
\cite{Dirac1973,Sen:1971,Utiyama:1973,Utiyama:1974,Freund:1974,Hayashi:1976,Hayashi:1978,Nishioka:1985,Ranganathan:1987,
Padmanabhan85,Mannheim89,Hochberg91,Wood92,Wheeler98,Feoli98,Pawlowski99,Nishino2009,Demir2004,Mannheim09,Huang1989,Wei2006}.
In the model considered in Refs. \cite{JMS,AJS,JM09,JM10_1,JM10_2}, scale
invariance is broken by the phenomenon which we call as cosmological symmetry
breaking \cite{JM,JMS}. Here, motivated by the Big Bang cosmology, we assume a
spatially flat background Friedmann-Robertson-Walker (FRW) metric. The model 
admits a classical time dependent solution which breaks scale invariance. 
This generates all the dimensionful parameters in the theory such as the Hubble 
parameter, the Planck mass, the vacuum or dark energy, Weyl meson mass as well 
as the electro-weak particle masses \cite{JMS,JM09,JM10_1,JM10_2}.

In this paper we consider a model where 
the matter sector displays global 
scale invariance. The gravitational term in the action, however, breaks 
scale invariance. Before electro-weak symmetry breaking, the matter sector is 
invariant under global scale transformation as it contains 
no dimensionful parameter. This requires the introduction of an additional 
real scalar 
field apart from the standard Higgs field. The scale invariance is ultimately 
broken spontaneously along with the electro-weak symmetry. The gravitational action is the standard
Hilbert-Einstein action with an explicit cosmological constant 
and hence violates scale invariance.  
For the model considered in Ref. 
\cite{JM10_2}, it has been shown that pure gauge fields lead to vanishing cosmological 
constant at all orders in the gauge field coupling. This follows for all the
gauge fields which do not have any coupling to the Higgs particle. Similar
result applies to fermion fields, as long as we ignore their coupling to Higgs. 
All of these results are applicable for the model considered in the present paper.
However the problem of fine tuning of the cosmological constant
was addressed only partially in Ref. \cite{JM10_2}
as we were not able to explicitly rule out
the possibility of large contributions from cosmological constant
from the matter sector. The basic problem arose due to the fact that
Higgs field was coupled non-minimally to gravity. However, no such problem
exists in the present model and we demonstrate that due to exact scaling symmetry 
in the matter action cosmological constant receives no contribution from the matter sector.

Scale invariance is generally believed to be anomalous. However it has been shown 
that scale transformations can be extended to arbitrary dimensions such that the 
action is invariant under this generalized transformation 
\cite{Englert:1976,JMS,JM09,JM10_1,JM10_2,Shaposhnikov:2008a,Shaposhnikov:2008b}.
Within dimensional regularization, the scale anomaly arises primarily since the action is not invariant
under scale transformations in dimensions other than four \cite{Delbourgo}.
In the present case the matter action is made to be scale invariant in arbitrary 
dimensions by introducing terms with fractional powers of the additional scalar 
field. 
Hence, as long as we neglect gravitational corrections, 
once the action is specified in $d$ dimensions, we can directly extract 
dimensionally regularized Feynman amplitudes for the matter sector\cite{Ramond}.
However, we note that this theory makes sense only if the scalar field, which is 
raised to a fractional power, is non-zero classically. In this case one may make 
an expansion around this classical value and the theory is well defined. 
However the theory is ill defined if the minimum of the potential
arises at zero values of the classical scalar fields. 
If, for 
some reason, one demands that only terms with integral powers have to appear in 
the action, then the theory necessarily has scale anomaly.

As an alternative, we also consider a model based on the idea
of unimodular gravity \cite{Weinberg,VanDer,Buchmuller,Buchmuller1,Henneaux,Henneaux89,Sorkin}. Here we suggest a model which does not respect
the full general coordinate transformations but only their unimodular
subset. The matter sector of this theory is invariant under pseudoscale
transformations \cite{ChengPRL,JM10_1} but not under the standard scale
transformations. We show that this is sufficient to constrain the 
contribution to the cosmological constant from the matter sector.

The present paper is organized as follows. In section \ref{sec:realscalar} we 
describe the model and discuss the allowed ranges of various parameters of the
model. In section \ref{sec:CC} we compute the contribution of the matter sector
to the cosmological constant. In section \ref{sec:EXP} we expand around the 
classical solution to display the particle spectrum of this theory. In section
\ref{sec:UMG} we discuss the alternative model based on the idea
of unimodular gravity and finally in section \ref{sec:concl} we conclude.

\section{A Model with Scale Invariant Matter Action}\label{sec:realscalar}
Consider the following action in $d=4-\epsilon$ dimensions
containing gravity, a scalar field $\Phi$ and the
standard model fields,
\ba
\mathcal{S} &=& \int d^dx \sqrt{-  g}\Bigg[{M_{\rm PL}^2\over 16\pi}R
-\Lambda + {\cal L}_{\rm Matter} \Bigg],
\label{eq:model1}
\ea
where $\Lambda$ is the cosmological constant, $M_{\rm Pl}$ is the Planck's
constant and 
 ${\cal L}_{\rm Matter}$ refers to the
matter Lagrangian which includes the standard model fields and the additional
real scalar field $\Phi$.
We choose the conventions followed in Refs.
\cite{Donoghue1,Donoghue2} where
the flat space-time metric takes the form $(1,-1,-1,-1)$.
The curvature tensor and its contractions are defined as,
\ba
 R^\mu_{\nu\alpha\beta}&=& -\partial_{\beta}\Gamma^\mu_{\nu\alpha} +\partial_{\alpha}\Gamma^\mu_{\nu\beta} +\Gamma^\mu_{\gamma\alpha}\Gamma^\gamma_{\nu\beta}
 -\Gamma^\mu_{\gamma\beta}\Gamma^\gamma_{\nu\alpha},\nn\\
R_{\nu\beta} &=& R^\mu_{\nu\beta\mu} \hspace{1mm} ,\hspace{4mm} R = R_{\nu\beta}g^{\nu\beta}.
\label{eq:R_notation}
\ea
The matter Lagrangian is given by,
\ba
{\cal L}_{\rm Matter} &=&  g^{\mu\nu} (D_\mu {\mc H})^\dagger (D_\nu{\mc H})
+ {1\over 2}  g^{\mu\nu} \partial_\mu {\Phi } \partial_\nu{\Phi} -
{\lambda_1\over 4}
(\mc H^\dagger\mc H- \lambda_2\Phi^2)^2(\Phi^2)^{-\delta}\cr
&-& \frac14
 g^{\mu\nu} g^{\alpha\beta}(\mathcal{A}^i_{\mu\alpha} \mathcal{A}^i_{\nu\beta}
+ \mathcal{B}_{\mu\alpha} \mathcal{B}_{\nu\beta} + \mathcal{G}^j_{\mu\alpha}
 \mathcal{G}^j_{\nu\beta}) (\Phi^2)^{\delta}
 + {\cal L}_{\rm fermions},
\label{eq:S_matter}
\ea
where $\delta=(d-4)/(d-2)$,
$\mc{H}$ is the Higgs doublet, $\mc{G}^j_{\mu\nu}$, $\mc{A}^i_{\mu\nu}$ and
$\mc{B}_{\mu\nu}$
represent the field strength tensors for the $SU(3)$, $SU(2)$ and $U(1)$
gauge fields of the standard model respectively. The superscripts $i$ and $j$ on $\mc{A}^i_{\mu\nu}$
and $\mc{G}^j_{\mu\nu}$ represent the $SU(2)$ and $SU(3)$ indices respectively.
All the repeated indices are implicitly summed over.
The scalar potential of this model is identical to what is proposed in
Ref. \cite{Shaposhnikov:2008a}.
The fermionic Lagrangian in d-dimensions is given by,
\ba
{\cal L}_{\rm fermions} =\left({\overline\Psi}_{\rm L}i
\gamma^\mu  {\cal D_\mu} \Psi_{\rm L} +
{\overline\Psi}_{\rm R}i \gamma^\mu  {\cal D_\mu} \Psi_{\rm R}  \right)
- \left[g_{_Y}\overline{\Psi}_{\rm L} {\mc H}\Psi_{\rm R} (\Phi^2)^{-\delta/2} + h.c.\right],
\label{eq:fermionLag}
\ea
where $\gamma^\mu=e^\mu_{~a}\gamma^a$,
$e_\mu^a$ is the vierbein field
and $a, b$ are Lorentz indices.
Here $\Psi_{\rm L}$ is an $SU(2)$ doublet, $\Psi_{\rm R}$ a singlet and
$g_{_Y}$ represents
the Yukawa coupling for the field $\Psi$. For simplicity we have displayed only one Yukawa
coupling term and included only
one family. In Eq. \ref{eq:fermionLag}, $D_\mu$ represents the covariant derivative of the fermion field,
\begin{equation}
{\cal D}_\mu \Psi_{\rm L/R} = \left({\tilde D}_\mu + {1\over2}\omega_{\mu}^{ab}\sigma_{ab}\right)\Psi_{\rm L/R}\ ,
\label{eq:Dfermion}
\end{equation}
where ${\tilde D}_{\mu}\Psi_{\rm L} = \partial_\mu -i g{\bf T\cdot A}_\mu -
ig'(Y^{\rm L}_{f}/ 2) B_\mu$,
${\tilde D}_{\mu}\Psi_{\rm R} = \partial_\mu - ig'(Y^{\rm R}_{f}/ 2) B_\mu$,
$\sigma_{ab} = {1\over4}[\gamma_a,\gamma_b]$ and $\omega_\mu^{ab}$ represents
the spin connection. 
Here $A_\mu$ is the $SU(2)$ field, $B_\mu$ the $U(1)$ field, ${\bf T}$ represents
the SU(2) generators and $Y_f$ the U(1) hypercharges. The color interaction
of quarks has been suppressed for simplicity.

The terms such as $(\Phi^2)^\delta$ which involve the scalar field raised
to a fractional power are handled by making a Taylor expansion around the 
classical value of the field \cite{Englert:1976,JMS,JM09,JM10_1,JM10_2,Shaposhnikov:2008a,Shaposhnikov:2008b}. Hence these make sense only if the classical
value is non-zero. If the minimum of the potential arises at zero value 
of the classical field then the model is ill defined. If the classical
value of the scalar field $\Phi$ is non-zero then these terms provide
an infinite series of terms involving higher and higher powers of
the fluctuating fields. These higher order terms might cause some complications
at loop orders, as discussed below and in Refs. 
\cite{Shaposhnikov:2008a,Shaposhnikov09}. 

The scaling symmetry of the action in Eq. \ref{eq:model1} is spoiled by the
presence of the dimensionful parameters, 
 $M_{\rm PL}$ and $\Lambda$, in the gravitational
sector of the model. The cosmological constant may, in principle, also 
acquire contributions at loop orders from the matter sector.  
However, as we shall show, the matter sector gives
identically zero contribution to the vacuum energy and hence to the cosmological
constant since the corresponding action respects scale invariance.
 We point out that due
to the presence of the terms with fractional powers of the scalar field $\Phi$ it displays
global scale invariance in arbitrary dimensions. Hence within the framework
of dimensional regularization, all the regularized 
Feynman amplitudes respect scale invariance. 

In our model both the scalar fields
acquire non zero classical solutions and the scaling symmetry in the matter action
is broken spontaneously. The potential is minimized if 
\cite{Shaposhnikov:2008a},
\ba
\mc H^\dagger\mc H= \lambda_2\Phi^2.
\label{eq:Higgs}
\ea
We assume that classically,
\cite{Shaposhnikov:2008a},
\ba
\Phi = \Phi_0,
\label{eq:Phi_0}
\ea
where $\Phi_0$ is a dimensionful parameter.
We point out that Eqs. \ref{eq:Higgs} and \ref{eq:Phi_0} solve 
the classical equations of motion.
We denote the classical Higgs field by $\mc H_0$, {\it i.e.},
\begin{equation}
\mc H_0 = {1\over \sqrt{2}}\left(\matrix{0\cr v} \right),
\label{eq:Higg0}
\end{equation}
where $v$ is the electro-weak symmetry breaking scale.

In $d=4$, the extra scalar field $\Phi$ couples directly only to the Higgs field but
 in any other dimension it couples to other matter fields via the fractional powers
of $\Phi$ in different terms of the matter Lagrangian. Hence it may contribute at loop orders.
As in earlier papers \cite{Englert:1976,JMS,JM09,JM10_1,JM10_2,Shaposhnikov:2008a,Shaposhnikov:2008b},
the scalar field terms raised to fractional powers, such as
$(\Phi^2)^{-\delta}$, are handled by making an expansion around
their classical values,
\ba
\Phi &=& \Phi_0 + \hat\Phi.
\label{eq:Phichi}
\ea
For any process the higher order terms in this expansion give
contributions at least of order $p^2/\Phi_0^2$ where $p$ is the momentum
scale of the process under consideration. Hence the theory is
perturbatively reliable only up to the scale $\Phi_0$ \cite{Shaposhnikov09}.
The parameter $\Phi_0$ has to be chosen suitably so that perturbation
theory is valid for sufficiently large mass scale compared to the electro-weak scale.
As the two parameters, $\lambda_2$ and $\Phi_0$, are related to the electro-weak scale as,
\ba
v^2=2\lambda_2\Phi_0^2,
\ea
for large $\Phi_0$, the parameter, $\lambda_2$ has to be very small to generate the
correct electro-weak scale. It has been shown in Ref. \cite{Shaposhnikov:2008a} that
a small value of $\lambda_2$ does not lead to
any fine tuning problem. Hence for $\Phi_0$ sufficiently larger than the electro-weak scale,
the model is free of fine tuning and is consistent with present observations.
One may, of course, consider a supersymmetric generalization of this model which may 
provide an alternate mechanism to stabilize the scalar potential.

Although in general scale invariance
would allow terms with three independent parameters in the potential term of the scalar fields,
we have included only two. The reason for this particular choice of potential term
is discussed in detail in Ref. \cite{Shaposhnikov:2008a}, where
the authors obtain the one loop effective action corresponding to this model.
If we include all the possible terms then the potential does
not have a minimum at nonzero values of the scalar fields. In this case
the theory is not phenomenologically acceptable and is also theoretically
ill defined due to the fractional powers of the field $\Phi$. 
However, the fact that one of the allowed terms is not present
in the action implies that we may not be able to absorb all the divergences
into renormalization of the parameters of the model. This term will
be generated at loop orders. In order to cancel this we may require
introduction of additional terms involving the scalar fields, 
spoiling the renormalization of the model.
The renormalization of the model may also be spoiled by higher order terms
in the expansion of terms such as $(\Phi^2)^{-\delta}$. However even if
the theory is not renormalizable, phenomenologically it is not a serious problem
\cite{Shaposhnikov:2008a,Shaposhnikov09},
since the theory is perturbatively reliable only at low
energies, $E\ll \Phi_0$,  where the additional contributions are very small.
Furthermore we point out that it is
possible to introduce additional terms in the action which vanish in the limit
$d=4$. For example, in full generality we may replace,
\begin{equation}
(\Phi^2)^{-\delta} \rightarrow [\Phi^2 + \lambda_3 \mc H^\dagger \mc H]^{-\delta}
\end{equation}
in Eq. \ref{eq:S_matter}. This introduces an additional parameter $\lambda_3$
which might be useful in canceling some of the divergences in the scalar
field sector of this theory. The entire issue of renormalizability of
these models is more complicated and we postpone it to further research.

Before we proceed to compute the contribution of the matter sector to the cosmological
constant a comment on the gravitational action is
in order. Although the gravitational action breaks scale invariance in this model,
it is possible to generalize this action to make it scale invariant by introducing
an additional scalar field and setting $\Lambda=0$. In this case scale invariance in the
gravitational sector could be broken
by a mechanism we call cosmological symmetry breaking \cite{JMS,AJS,JM09,JM10_1,JM10_2}
where a classical time dependent solution breaks the scale invariance. 
Assuming an FRW metric, we find an exact solution with the
scale parameter expanding exponentially with time \cite{JMS}. In this case the Planck
constant as well as the Hubble constant are generated dynamically \cite{JMS}.
The theory also generates an effective cosmological constant. This mechanism
has some similarity to the phenomenon of phase transitions induced by
background curvature \cite{Shore,Cooper,Allen,Ishikawa,Buchbinder85,Wetterich1,Perez,Finelli,Odintsov:1990,Odintsov:1991,Buchbinder,Odintsov:1993,Wetterich2}
but is not identical. A theory where the entire action is scale invariant
is aesthetically appealing. However the action displayed in Eq. \ref{eq:model1}
is simpler to handle and closer to the Einstein's gravity. We postpone
a detailed analysis of the scale invariant gravitational action to a
future paper.

\section{Contribution to the Cosmological Constant}\label{sec:CC}
The important point about this model is that the matter Lagrangian
displays exact scale invariance in arbitrary dimensions. The Lagrangian of
Eq. \ref{eq:S_matter} will directly lead to dimensionally regulated amplitudes.
As the scale current is conserved in arbitrary dimensions,
we do not expect any scale anomaly
\cite{Englert:1976,JMS,JM09,JM10_1,JM10_2,Shaposhnikov:2008a,Shaposhnikov:2008b}.
This implies that the trace of the energy momentum tensor must
vanish and there would be no contribution to the cosmological constant
from the matter part of the action. We next verify this explicitly.

The energy momentum tensor following from the $d$ dimensional action may
be expressed as,
\ba
T_{\alpha\beta} &=& -g_{\alpha\beta}{\mc L}
+(D_\alpha{\mc H})^\dagger (D_\beta{\mc H})
+(D_\beta{\mc H})^\dagger (D_\alpha{\mc H})\cr
&&+{1\over 2} \partial_\alpha \Phi \partial_\beta\Phi
+{1\over 2} \partial_\beta \Phi \partial_\alpha\Phi
- g^{\mu\nu} B_{\alpha\mu} B_{\beta\nu} (\Phi^2)^\delta\cr
&&+ \frac12\bar\Psi_{\rm L} i\gamma_\alpha D_\beta\Psi_{\rm L}
+ \frac12\bar\Psi_{\rm R} i\gamma_\alpha D_\beta\Psi_{\rm R}
+ \frac12\bar\Psi_{\rm L} i\gamma_\bt D_\al\Psi_{\rm L}
+ \frac12\bar\Psi_{\rm R} i\gamma_\bt D_\al\Psi_{\rm R},
\label{eq:emtensor}
\ea
where we have explicitly displayed the contribution coming from only one vector field.
The other vector fields give similar contributions. The equations of motion for the scalar
fields are given by,
\ba
D^\mu D_\mu {\mc H} &=& - {\lambda_1\over 2} ({\mc H}^\dagger {\mc H} - \lambda_2\Phi^2){\mc H} (\Phi^2)^{-\delta} -g_{_Y} \bar \Psi_{\rm R}
\Psi_{\rm L}
(\Phi^2)^{-\delta/2},
\label{eq:higgseom}\\
\Phi\partial^\mu\partial_\mu\Phi &=& \lambda_1\lambda_2 ({\mc H}^\dagger {\mc H} - \lambda_2\Phi^2) \Phi^2 (\Phi^2)^{-\delta} + {\lambda_1\delta\over 2}
({\mc H}^\dagger {\mc H} - \lambda_2\Phi^2)^2(\Phi^2)^{-\delta}\cr
&&- {\delta\over 2} B^{\alpha\mu} B_{\alpha\mu} (\Phi^2)^{\delta}
+ \delta\, g_{_Y}\left[\overline{\Psi}_{\rm L} {\mc H}\Psi_{\rm R} (\Phi^2)^{-\delta/2} + h.c.\right]
\label{eq:Phieom}
\ea
and the equations of motion for the fermion fields may be written as,
\ba
i\gamma^\mu D_\mu\Psi_{\rm R} - g_{_Y}\mc H^{\dag}\Psi_{\rm L} (\Phi^2)^{-\delta/2} &=& 0,\\
i\gamma^\mu D_\mu\Psi_{\rm L} - g_{_Y}\mc H \Psi_{\rm R} (\Phi^2)^{-\delta/2} &=& 0.
\ea
Using the fermion equations of
motion, the trace of energy momentum tensor may be expressed as,
\ba
T^\alpha_{\alpha} &=& {1\over 2}(d-2) \left[{\mc H}^\dagger D^\alpha D_\alpha{\mc H} + h.c.\right]
+{1\over 2}(d-2) \Phi\partial^\alpha \partial_\alpha\Phi
- B^{\alpha\mu} B_{\alpha\mu} (\Phi^2)^\delta \cr
&+& d\left[{\lambda_1\over 4} ({\mc H}^\dagger {\mc H} - \lambda_2\Phi^2)^2
(\Phi^2)^{-\delta}+{1\over 4} B^{\alpha\mu} B_{\alpha\mu} (\Phi^2)^\delta  \right]\cr
&+& g_{_Y}\left[\overline{\Psi}_{\rm L} {\mc H}\Psi_{\rm R} (\Phi^2)^{-\delta/2} + h.c.\right],
\label{eq:emtrace}
\ea
up to terms which are total derivatives. These will drop out after integration
by parts and hence won't contribute to the cosmological constant.
Now using the equations of motion for the scalar fields
one finds that $T^\alpha_\alpha=0$. Hence the matter sector
of the theory gives zero contribution to the cosmological constant.
The entire contribution to the cosmological constant comes from the
gravitational sector of theory and hence no fine tuning may be required in
fitting its value to observations. 

We emphasize that Eq. \ref{eq:emtrace} lists all the operators that can
contribute to the trace of the energy momentum tensor in this theory
after dimensional regularization. Hence the cosmological constant in the
full quantum theory must vanish if the right hand side of Eq. \ref{eq:emtrace}
vanishes. In our demonstration
that $T^\alpha_\alpha=0$, we have used the classical
equations of motion. However this is perfectly justified since these are
valid as Heisenberg operator equations \cite{Nishijima}.

Apart from the terms in the Lagrangian in Eqs. \ref{eq:model1}, \ref{eq:S_matter} 
and \ref{eq:fermionLag} we are allowed to introduce
additional terms in the action such as, $\Phi^2 R$, ${\mc H}^\dagger
{\mc H}R$ without violating the scaling symmetry. We assume that
the couplings corresponding to these terms are negligible. Even at loop orders such
couplings will be generated only by quantum gravity effects and hence will
be suppressed.

The gravitational part of the action explicitly violates
scale invariance. It may contribute to the cosmological constant at
loop orders. The one loop
and two loop contributions due to the gravitation have been considered
earlier \cite{Mueller,Sawhill}. However the precise result for the 
gravitational contribution to cosmological constant is unknown due to
the usual problems in handling quantum gravity. Here we assume
that these contributions are small and give only small corrections
to the cosmological constant parameter introduced in our model.

\section{Expansion Around the Classical Solution}\label{sec:EXP}
The one loop effective
action for this model has already been computed in Ref.
\cite{Shaposhnikov:2008a}.
Here we briefly consider the quadratic terms in the expansion around
the classical solution, shown in Eqs. \ref{eq:Phi_0} and \ref{eq:Higg0}, in order to
display the particle spectrum of this theory. We have,
\begin{eqnarray}
\Phi = \Phi_0 + \hat \Phi ,\quad
\mc H = \mc H_0 + \hat{\mc H}.
\end{eqnarray}
We parametrize $\hat \mc H$ as,
\begin{equation}
\hat\mc H = {1\over \sqrt{2}}\left(\matrix{\xi_1+i\xi_2\cr \xi_3+i\xi_4}
\right)
\end{equation}
and denote the real Higgs field as $\hat\mc H_3 = v + \xi_3$. The electro-weak scale $v^2=2\lambda_2
\Phi_0^2$. We next expand the potential in $d$ dimensions around this
classical solution. We have,
\begin{eqnarray}
V &=& {\lambda_1\over 4}\left(\mc H^\dagger \mc H - \lambda_2\Phi^2\right)^2
(\Phi^2)^{-\delta} \cr
& = & {\lambda_1\over 4} (\Phi_0^2)^{-\delta} \left[(v\xi_3 - 2\lambda_2\Phi_0\hat\Phi)^2 + \ldots\right],
\label{eq:Potential}
\end{eqnarray}
where we display only the quadratic terms. The fields $\xi_3$ and $\hat\Phi$
mix with each other. We make an orthogonal transformation to diagonalize
the mass matrix. The transformed fields are given by,
\begin{eqnarray}
\phi_1 &=& \cos\theta \xi_3 - \sin\theta \hat\Phi,\cr
\phi_2 &=& \sin\theta \xi_3 + \cos\theta \hat\Phi,
\end{eqnarray}
where
\begin{equation}
\sin\theta = \sqrt{2\lambda_2 \over 1+ 2\lambda_2} \approx 0.
\end{equation}
From Eq. \ref{eq:Potential} we find that the field $\phi_1$ has mass
$m_1^2 = \lambda_1 v^2(1+2\lambda_2)/2$. The field $\phi_2$ acts as the
Goldstone boson corresponding to spontaneous breakdown of scale symmetry
with mass $m_2 = 0$.

The field $\phi_2$ is essentially the field $\hat \Phi$. It couples
indirectly to the matter fields through the Yukawa interaction terms.
The coupling at tree level is zero. However we get a coupling proportional
to the mass of matter particles at loop orders. Since this particle
has zero mass it will have long range interactions and hence can lead to
a modification in Newtonian gravity. The precise value of the correction
depends on $\Phi_0$. If $\Phi_0$ is comparable or much larger than the
Planck mass then the correction may be small and negligible. In the present
paper we do not consider the phenomenological implications of this
particle. However it is clear that for sufficiently large value of
$\Phi_0$ its contribution will not violate the observational bounds.

\section{Unimodular Gravity}\label{sec:UMG}
In this section we shall slightly deviate from the main development of this paper and 
suggest an alternative model based on the idea
of unimodular gravity \cite{Weinberg,VanDer,Buchmuller,Buchmuller1,Henneaux,Henneaux89,Sorkin}.
Motivated by the work of Refs. \cite{Zee,Buchmuller,Buchmuller1} and
the discussion in Ref. \cite{Weinberg} we consider the possibility
that the quantum theory may not respect the full general coordinate
invariance, $x^\mu\rightarrow x'^\mu$. 
It may respect only the restricted invariance where
$det(\partial x'^\mu/\partial x^\nu)=1$. However the classical theory
respects the full
general coordinate transformations. In $d$ dimensions the matter action 
in this case may be obtained by replacing $(\Phi^2)^{\dl}$ by 
$(\sqrt{-g}\,)^{(4-d)/d}$ in Eq. \ref{eq:S_matter} as,
\ba
{\cal L}_{\rm Matter} &=&  g^{\mu\nu} (D_\mu {\mc H})^\dagger (D_\nu{\mc H})
+ {1\over 2}  g^{\mu\nu} \partial_\mu {\Phi } \partial_\nu{\Phi} -
{\lambda_1\over 4}
(\mc H^\dagger\mc H- \lambda_2\Phi^2)^2(\sqrt{-g}\,)^{(d-4)/d}\cr
&-& \frac14
 g^{\mu\nu} g^{\alpha\beta}(\mathcal{A}^i_{\mu\alpha} \mathcal{A}^i_{\nu\beta}
+ \mathcal{B}_{\mu\alpha} \mathcal{B}_{\nu\beta} + \mathcal{G}^j_{\mu\alpha}
 \mathcal{G}^j_{\nu\beta}) (\sqrt{-g}\,)^{(4-d)/d}
 + {\cal L}_{\rm fermions},
\label{eq:S_matter_um}
\ea
with the fermion Lagrangian similarly modified. The dimensionally regulated
Feynman amplitudes obtained from this Lagrangian will not obey the full
general coordinate invariance. They will only obey the limited transformation 
rule under which $det(\partial x'^\mu/\partial x^\nu)=1$.
They will also not obey scale invariance.
Hence scale transformations are anomalous in the present model. However
the model is invariant under pseudoscale transformations in $d$ 
dimensions \cite{ChengPRL,JM10_1}. Hence the dimensionally regulated Feynman amplitudes would
obey this symmetry as long as we consider only the matter sector of
the theory and neglect the gravitational corrections to the matter sector. 

For completeness, here we review the pseudoscale transformations 
\cite{ChengPRL,JM10_1}. In $d$
dimensions the transformation law is as follows,
\ba
x & \rightarrow & x\, , \cr
\Phi &\rightarrow & \Phi/\Omega\, ,\cr
{ g}^{\mu\nu} &\rightarrow & { g}^{\mu\nu}/\Omega^{b(d)}\, ,\cr
{ g}_{\mu\nu} &\rightarrow & { g}_{\mu\nu}\Omega^{b(d)}\, ,\cr
A_\mu &\rightarrow& A_\mu\, ,\cr
\Psi &\rightarrow& \Psi/\Omega^{c(d)},
\label{eq:pseudo_ddim}
\ea
where $b(d) = 4/(d-2)$ and $c(d) = (d-1)/(d-2)$. In Eq. \ref{eq:pseudo_ddim}
$\Phi$ represents any scalar field, $A_\mu$ a vector field and $\Psi$ a
spinor field. The action corresponding to the Lagrangian given
in Eq. \ref{eq:S_matter_um}
is invariant under this transformation in $d$ dimensions. Hence
the dimensionally regulated Feynman amplitudes would preserve this symmetry.
However the usual scale transformations, under which $g_{\mu\nu}$ does not
change but $x\rightarrow \Omega x$ are anomalous. Hence the coupling
$\lambda_1$ would be dimensionful in dimensions $d$ different from 4, 
indicating breakdown
of scale invariance. Similarly the gauge coupling would be dimensionful
in dimensions $d$ different from 4. 

The cosmological constant or vacuum energy term, however,
is not invariant under pseudoscale transformations. Hence we do not require
the full scale transformations to constrain the vacuum energy contributions
from matter sector. Pseudoscale transformations are sufficient for this
purpose. 

 It is clear that the modified theory is 
consistent with all observations. Classically, in four dimensions, it produces
same results as Einstein's gravity. It's predictions differ only when we
consider quantum gravity effects which are very tiny at the energy scales
probed by the current experiments.
We point out that in contrast to the model described in Eq. \ref{eq:S_matter}, the
present model is valid perturbatively when we expand around the trivial
classical solution where the background classical
fields are zero.

The matter sector of the model shows exact pseudoscale invariance \cite{ChengPRL, JMS,JM10_1, JM10_2} in the
full quantum theory. As discussed above the cosmological constant term 
violates pseudo-scale invariance. Hence, in analogy
with the calculation in Section \ref{sec:CC}, we again expect to find zero
contribution to the cosmological constant from
the matter fields. This may also be verified explicitly. We have checked that
the matter action produces vanishing
$T^\mu_\mu$ in $d$ dimensions, exactly as was found in Section 3 for the
model discussed in Section 2. Hence the matter fields produce
zero vacuum energy at all orders in perturbation theory, as long as we
ignore corrections due to gravitational coupling. 

The main phenomenological advantage of the action, Eq. \ref{eq:S_matter_um},
is that the couplings of
$\Phi$ to the Standard Model fields are now extremely tiny. They will arise
only due to its mixing with the Higgs boson which is proportional to
$\lambda_2$. Hence in this case we may allow a much wider range of values for
the scale $\Phi_0$.

\section{Conclusions}\label{sec:concl}
In this paper we have presented a generalization of the standard model of particle
physics which displays exact quantum scale invariance except in the gravitational
sector. We explicitly show that within the framework of dimensional
regularization, the matter sector of the theory gives identically zero
contribution to the cosmological constant. This contribution vanishes
due to scale or pseudoscale 
invariance and applies to the full quantum theory. The
entire contribution to the cosmological constant arises from the gravitational
sector of theory. Hence the loop corrections are naturally suppressed
and no fine tuning may be required to fit the observations. This shows that
scale or pseudoscale invariance may provide a natural solution to the problem of cosmological
constant.

\bigskip
\noindent
{\bf \large Acknowledgements:}  Naveen Kumar Singh thanks the Council
of Scientific and Industrial Research (CSIR), India for providing his Ph.D.
fellowship. His fellowship number is  F.No.09/092(0437)/2005-
EMR-I.

\end{spacing}
\begin{spacing}{1}
\begin{small}

\end{small}
\end{spacing}
\end{document}